\documentclass[11pt]{article}
\usepackage{moriond,epsfig}

\def\be{\begin{equation}}
\def\ee{\end{equation}}
\def\bea{\begin{eqnarray}}
\def\eea{\end{eqnarray}}

\usepackage{amsmath}
\usepackage{ifthen}
\newboolean{uprightparticles}
\setboolean{uprightparticles}{false} 
\newboolean{articletitles}
\setboolean{articletitles}{true} 
\usepackage{amssymb}
\usepackage{amsfonts}
\usepackage{upgreek}
\usepackage{xspace}
\usepackage{lhcb-symbols-def}
\usepackage{hyperref}
\usepackage{url}
\usepackage{cite}
\usepackage{mciteplus}

\begin{document}
\vspace*{4cm}


\title{Results on direct \CP violation in $B$ decays in LHCb}

\author{ T.M. Karbach \footnote{on behalf of the LHCb collaboration} }

\address{Technische Universit\"at Dortmund, Dortmund, Germany}

\maketitle\abstracts{
I present three studies from the LHCb experiment on the subject of direct \CP
violation in \Bz and \Bs decays.
First, we measure the \CP asymmetry in $\Bpm\to\psi\Kpm$
decays, with $\psi=\jpsi, \psi(2S)$, using $0.35\invfb$ of $pp$ collisions at
$\sqrt{s}=7\tev$. We find no evidence for \CP violation.
Second, using the same data sample, we see the first evidence of \CP 
violation in the decays of \Bs mesons to $\Kpm\pimp$ pairs, 
$A_{\CP}(\Bs\to K\pi) = 0.27 \pm 0.08\stat\pm 0.02\syst$ ($3.3\sigma$). 
%
Third, using $1.0\invfb$ of data, measurements of \CP sensitive observables 
of the $\Bpm\to D\Kpm$ system are presented. They include the first observation 
of the suppressed mode $\Bpm\to[\pipm\Kmp]_D\Kpm$. Combining several $D$ final states, \CP 
violation in $\Bpm\to D\Kpm$ decays is observed with a significance of $5.8\sigma$.
}


\section{Measurement of \CP asymmetries in $\Bpm\to\psi h^\pm$ decays}

\begin{figure}
\centering
\includegraphics[width=.9\textwidth]{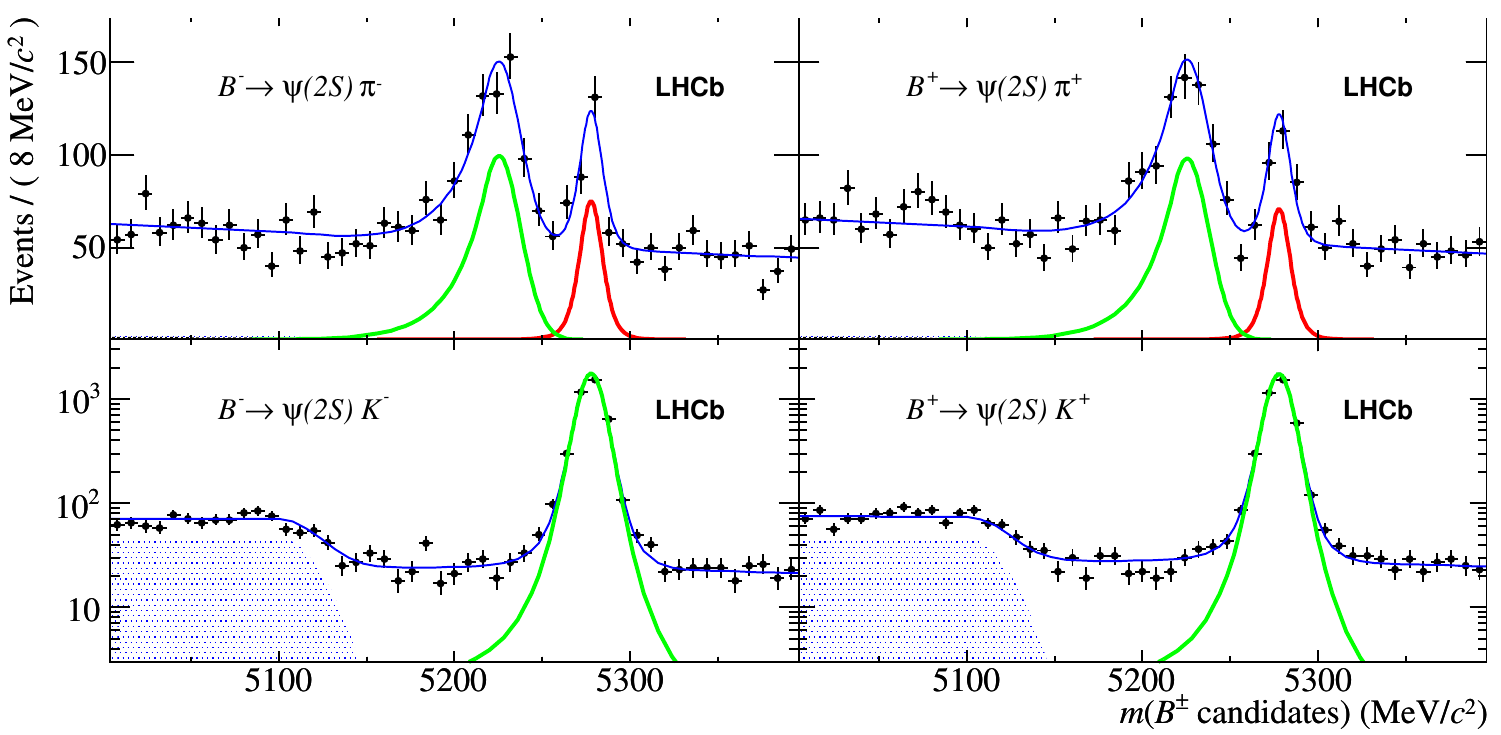}
\caption{$\Bpm\to\Psi(2S) h^\pm$ invariant mass distributions, overlaid by the total fitted 
PDF (thin line). Pion-like events are reconstructed as $\jpsi\pipm$ and 
enter in the top plots. All other events are reconstructed as $\jpsi\Kpm$ enter the 
bottom plots (shown in logarithmic scale). \Bm decays are shown on the left, \Bp on the right. The 
dark [red] curve shows the $\Bpm\to\psi(2S)\pipm$ component, the light [green] curve represents
$\Bpm\to\psi(2S)\Kpm$. Partially reconstructed backgrounds are shaded.
\label{fig:1-3}}
\end{figure}

The $\Bpm\to\psi h^\pm$ decays, with $\psi = (\jpsi, \psi(2S))$ and $h=K, \pi$, receive contributions
from both tree and penguin diagrams. If these contributions have different weak phases, direct \CP
violation may occur. The Standard 
Model predicts that for $b\to c\bar{c}s$ decays the tree and penguin contributions 
have the same weak phase and thus no direct \CP violation is expected in \mbox{$\Bpm\to\psi\Kpm$}.
For $b\to c\bar{c}d$ transitions, however, both contributions have different weak phases, and \CP 
violation in $\Bpm\to\psi\pipm$ decays may occur.
Their branching fractions are expected
to be about 5\% of the favoured $\Bpm\to\psi\Kpm$ modes. 
In our paper~\cite{Aaij:2012jw} we analyse a data sample
of $0.35\invfb$ of $pp$ collisions at $\sqrt{s}=7\tev$, taken in 2011 with the LHCb
detector.
We define the \CP asymmetry and the charge-averaged ratio of branching ratios as
\begin{equation}
  A^{\psi\pi} = \frac{\BR(\Bm\to\psi\pim)-\BR(\Bp\to\psi\pip)}{\BR(\Bm\to\psi\pim)+\BR(\Bp\to\psi\pip)}~,
  \quad
  R^{\psi} = \frac{\BR(\Bpm\to\psi\pipm)}{\BR(\Bpm\to\psi\Kpm)}~.
\end{equation}
The $\psi$ resonance is reconstructed in the $\mup\mu^-$ final state,
and the well known and abundant decay \mbox{$\Bpm\to\jpsi\Kpm$} is used as a control channel.
It is crucial to control its cross feed
into the $\Bp\to\jpsi\pip$ channel. Here we benefit from LHCb's two ring imaging
Cherenkov (RICH) detectors that provide strong $K/\pi$ separation.
%
We obtain the signal yields from a simultaneous fit to the $B$ candidate invariant
mass distribution in eight independent subsamples, defined by the charge ($\times 2$), the $\psi$
state ($\times 2$) and the flavour of the bachelor hadron ($K,\pi$, $\times 2$). The
fit projections for the $\psi(2S)$ subsamples are shown in Figure~\ref{fig:1-3}.
The measured ratios of branching fractions are 
$R^{\jpsi} = (3.83 \pm 0.11 \pm 0.07) \times 10^{-2}$ and
$R^{\psi(2S)} = (3.95 \pm 0.40 \pm 0.12) \times 10^{-2}$,
where the first uncertainty is statistical and the second systematic. $R^{\psi(2S)}$ is 
compatible with the one existing measurement~\cite{Bhardwaj:2008ee}, $(3.99 \pm 0.36 \pm 0.17)\times 10^{-2}$.
The measurement of $R^{\jpsi}$ is $3.2\sigma$ lower than the
current world average~\cite{PDG2010}, $(5.2 \pm 0.4) \times 10^{-2}$.
Using the established measurements of the Cabibbo-favoured branching 
fractions~\cite{PDG2010}, we deduce
$\BR(\Bpm \to \jpsi\pipm) = (3.88 \pm 0.11 \pm 0.15) \times 10^{-5}$, 
$\BR(\Bpm \to \psi(2S)\pipm) = (2.52 \pm 0.26 \pm 0.15) \times 10^{-5}$.
The measured \CP asymmetries,
\begin{align}
  A^{\jpsi\pi}_{\CP}    &= 0.005 \pm 0.027 \pm 0.011~, \\
  A^{\psi(2S)\pi}_{\CP} &= 0.048 \pm 0.090 \pm 0.011~, \\
  A^{\psi(2S)K}_{\CP}   &= 0.024 \pm 0.014 \pm 0.008~,
\end{align}
have comparable or better precision than previous results, and no evidence of 
direct \CP violation is seen.



\section{Direct \CP violation in $\Bz (\Bs) \to \Km\pip$ decays}

\begin{figure}
\centering
\includegraphics[width=.9\textwidth]{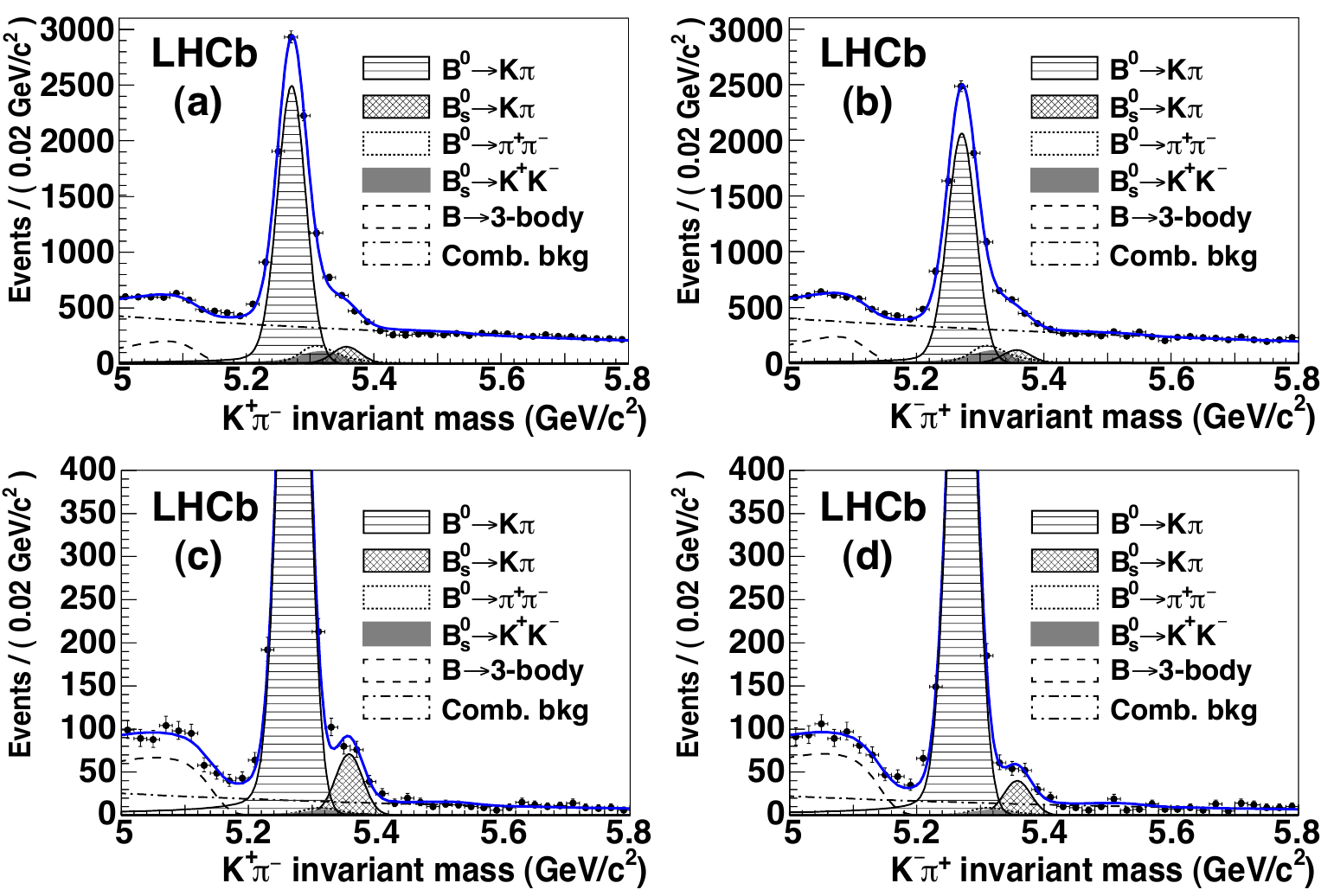}
\caption{$K\pi$ invariant mass spectra obtained using the event selection adopted 
for the best sensitivity on (a, b) $A_{\CP}(\Bz\to K\pi)$ and (c, d) $A_{\CP}(\Bs\to K\pi)$. 
Plots (a) and (c) represent the $\Kp\pim$ invariant mass whereas plots (b) and (d) represent 
the $\Km\pip$ invariant mass. The results of the unbinned maximum likelihood fits are overlaid. 
The main components contributing to the fit model are also shown.
\label{fig:2-1}}
\end{figure}

\CP violation is well established in the $K^0$ and $B^0$ meson systems. Recent 
results from LHCb have also provided evidence for \CP violation in the $D^0$ 
system~\cite{Aaij:2011in}. 
In our paper~\cite{Aaij:2012qe} we report evidence of direct \CP 
violation in the last neutral meson system, the $B^0_s$ system. We reconstruct
both $B^0 \rightarrow K^+\pi^-$ and $B^0_s \rightarrow K^- \pi^+$ decays 
in $0.35~\mathrm{fb}^{-1}$ of data collected with the LHCb detector in 2011.
The considered decays have contributions from both tree and penguin diagrams, and
are sensitive to contribution of new physics in the penguins. The \CP asymmetry in 
the $B^0 \rightarrow K^+\pi^-$ is well established~\cite{PDG2010}. The probability
or a $b$ quark to decay as $\Bs\to K\pi$ is about 14 times smaller than that to
decay as $\Bz\to K\pi$. However, both tree and penguin diagrams are roughly of the
same magnitude, so \CP violation effects can potentially be large.
We define the \CP asymmetries as
\begin{equation}
  A_{\CP}(B^0_{(s)}) = \frac{\Gamma(\overline{B}^0_{(s)} \to \bar{f}_{(s)}) 
                           - \Gamma(B^0_{(s)} \to f_{(s)})}
                            {\Gamma(\overline{B}^0_{(s)} \to \bar{f}_{(s)}
                           + \Gamma(B^0_{(s)} \to f_{(s)})}~,
\end{equation}
with $f=\Kp\pim$ and $f_s=\Km\pip$.
To distinguish the $\Kp\pim$ and $\Km\pip$ final states we rely on the RICH
particle identification system. We carefully control the efficiencies and
misidentification rates from data, through large control samples of
$D^*\to D\pi \to (K\pi)_D\pi$ and $\Lambda_b\to p\pi$ decays. There are cross feeds 
from $\Bz\to\pip\pim$ and $\Bs\to\Kp\Km$ decays, whose line shape we predict from simulation.
We compute a raw asymmetry from the yields of a fit to the invariant mass
distribution in the positive charge and negative charge subsamples. 
Figure~\ref{fig:2-1} shows the projections.
This raw asymmetry needs to be corrected for two effects: an inherent detector
charge asymmetry (which we estimate from our $D^*$ control samples) and a 
non-zero production asymmetry that is further diluted by $B$ mixing (thus it mostly affects
the \Bz channel due to its much slower $\Bz$--$\Bzb$ oscillation). The total
corrections to the raw asymmetry are $\Delta A_{\CP}(\Bz) = -0.007 \pm 0.006$
and $\Delta A_{\CP}(\Bs) = 0.010 \pm 0.002$, where the errors are statistical.
The systematic uncertainty of $A_{\CP}(\Bz)$ is dominated by uncertainties
due to instrumentation and production asymmetry, while the systematic uncertainty of $A_{\CP}(\Bs)$ 
receives a leading contribution from the combinatorial background description.
In conclusion we obtain the following measurements of the $C\!P$ asymmetries:
\begin{equation}
  A_{C\!P}(B^0 \rightarrow K\pi)=-0.088 \pm 0.011\,\mathrm{(stat)} \pm 0.008\,\mathrm{(syst)}\nonumber
\end{equation}
and
\begin{equation}
  A_{C\!P}(B^0_s \rightarrow K\pi)=0.27 \pm 0.08\, \mathrm{(stat)}\pm 0.02\,\mathrm{(syst)}.\nonumber
\end{equation}
The result for $A_{C\!P}(B^0 \rightarrow K\pi)$ constitutes the most precise 
measurement available to date. It is in good agreement with the current world 
average~\cite{HFAG}. The significance of the measured deviation from 
zero exceeds $6\sigma$. The result for $A_{C\!P}(B^0_s \rightarrow K\pi)$ is in 
agreement with the only measurement previously available~\cite{Aaltonen:2011qt}.
The significance computed for $A_{C\!P}(B^0_s \rightarrow K\pi)$ 
is 3.3$\sigma$, making this the first evidence for $C\!P$ violation in the decays 
of $B^0_s$ mesons. 

\section{Observation of \CP violation in $\Bpm \to D\Kpm$}

\def\ads{\ensuremath{ADS}\xspace}

\begin{figure}
\centering
\includegraphics[width=.9\textwidth]{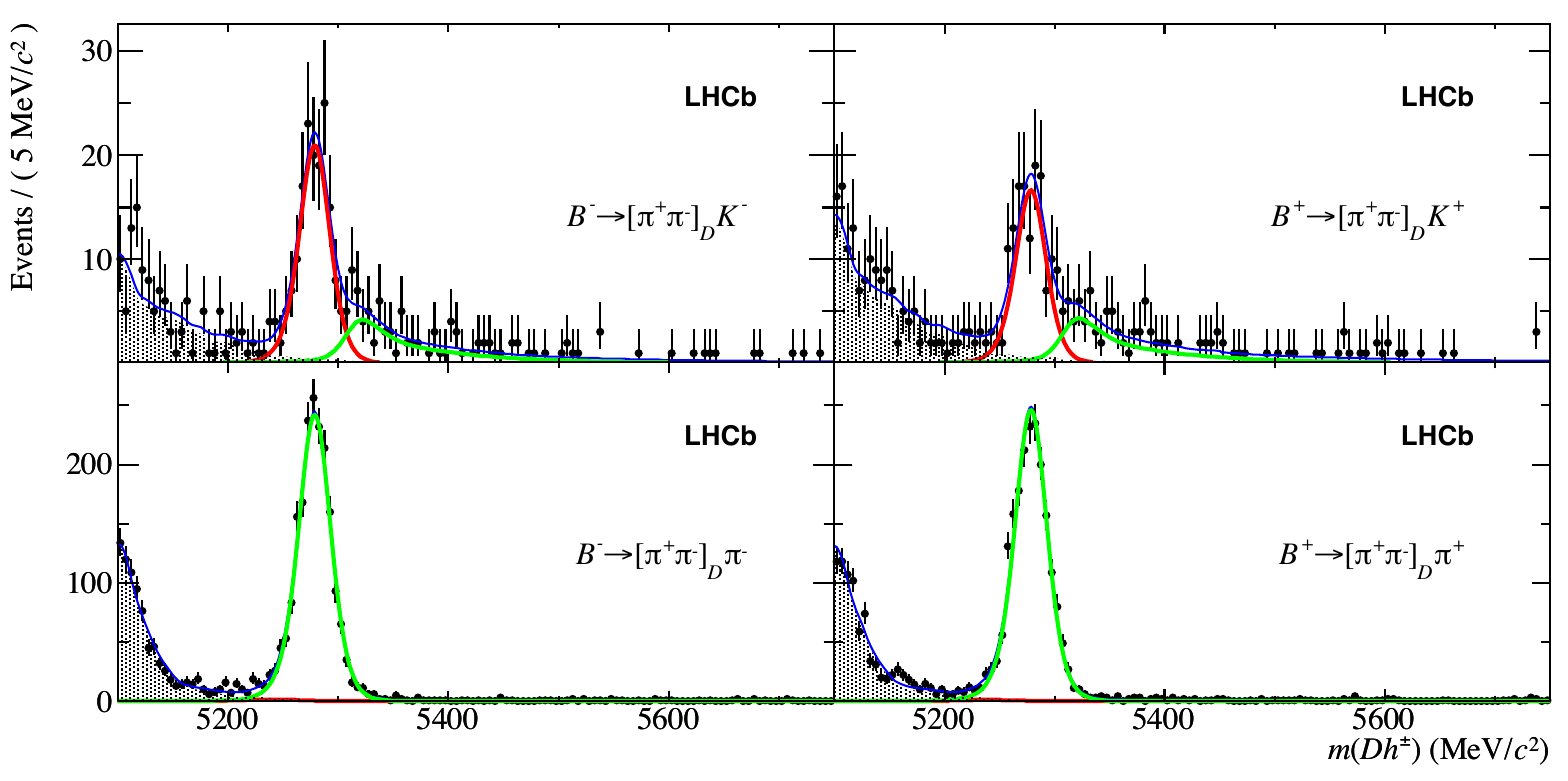}
\caption{The invariant mass distribution of selected $\Bpm\to[\pip\pim]_D h^\pm$ candidates.
The left plots are \Bm candidates, \Bp are on the right. In the top plots, 
the bachelor track passes the kaon RICH cut and the \B candidates are reconstructed 
assigning this track the kaon mass. The remaining events are placed in the bottom row 
and are reconstructed with a pion mass hypothesis. The dark (red) curve represents the 
$\B\to D\Kpm$ events, the light (green) curve is $\B\to D\pipm$. The shaded contribution 
are partially reconstructed events and the thin line shows the total PDF which also includes 
a linear combinatoric component.
\label{fig:3-3}}
\end{figure}

\begin{figure}
\centering
\includegraphics[width=.9\textwidth]{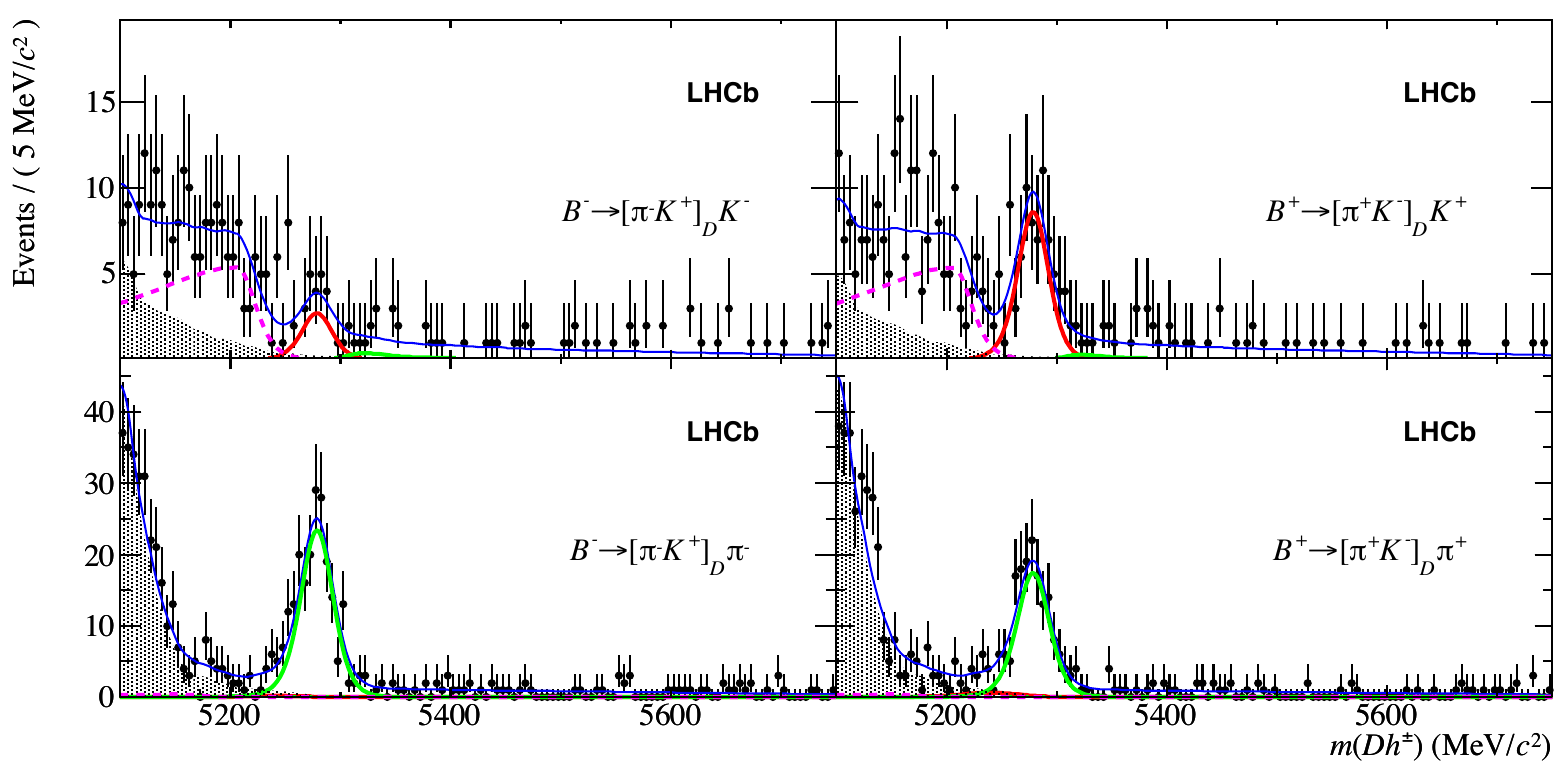}
\caption{The invariant mass distribution of selected $\Bpm\to[\pipm\Kpm]_D h^\pm$ candidates. 
See the caption of Fig.~\ref{fig:3-3} for a full description. The broken line here represents 
the partially reconstructed, but Cabibbo favoured, $\Bs\to\Dz\Kp\pim$ decays where the pion is 
lost.
\label{fig:3-4}}
\end{figure}

The CKM angle $\g=\arg\left(-V_{ud}V_{ub}^* / V_{cd}V_{cb}^* \right)$ is the
least well known angle of the corresponding unitarity triangle of the CKM
matrix $V$.
The angle \g can be measured in $\Bpm\to D\Kpm$ decays where the $D$ signifies 
a \Dz or \Dzb meson.
The amplitude for the $\Dz$ contribution is proportional to $V_{cb}$ whilst 
the $\Dzb$ amplitude depends on $V_{ub}$.
If the $D$ final state is accessible for both \Dz and \Dzb mesons, the two
amplitudes interfere and give rise to observables that are sensitive to \g.
Many different $D$ final states can be used.
In our analysis~\cite{Aaij:2012kz} of 1.0~\invfb of $\sqrt{s}=7~\tev$ 
data collected by \lhcb in 2011, we use the \CP eigenstates 
$D\to\Kp\Km$, $\pip\pim$ (often referred to as ``GLW'' modes~\cite{Gronau:1990ra,Gronau:1991dp}),
and the flavour eigenstate $D\to\pim\Kp$
(labelled ``ADS'' mode~\cite{Atwood:1996ci,Atwood:2000ck}).
The latter requires the favoured, $b\to c$ decay to be followed by a doubly Cabibbo-suppressed $D$ decay, 
and the suppressed $b\to u$ decay to be followed by a favoured $D$ decay.
As a consequence, the interfering amplitudes are of similar magnitude and 
hence large interference can occur.
In total, 13 observables are measured: three ratios of partial widths
\begin{equation}
  R_{K/\pi}^f = \frac{ \Gamma(\Bm\to [f]_D\Km)+\Gamma(\Bp\to [f]_D\Kp) }{ \Gamma(\Bm\to [f]_D\pim)+\Gamma(\Bp\to [f]_D\pip) }~,
  \label{eq:rkpi}
\end{equation}
where $f$ represents $KK$, $\pi\pi$ and the favoured $K\pi$ mode, six \CP asymmetries
\begin{equation}
  A_{h}^f = \frac{ \Gamma(\Bm\to [f]_Dh^-)-\Gamma(\Bp\to [f]_Dh^+) }{ \Gamma(\Bm\to [f]_Dh^-)+\Gamma(\Bp\to [f]_Dh^+) }~,
  \label{eq:acp}
\end{equation}
and four charge-separated partial widths of the \ads mode relative to the favoured mode
\begin{equation}
  R_h^{\pm}  = \frac{ \Gamma(\Bpm\to [\pipm\Kmp]_Dh^{\pm})}{ \Gamma(\Bpm\to  [\Kpm\pimp]_Dh^{\pm})}~.
  \label{eq:rh}
\end{equation}
Similar analyses have found evidence of the $\Bpm\to [\pipm\Kmp]_D\Kpm$
decay~\cite{Belle:2011ac,delAmoSanchez:2010dz,Aaltonen:2011uu}.
The abundant $\Bm\to D\pim$ decays have limited sensitivity to \g and provide a 
large control sample from which probability density functions are shaped.
The analysis method benefits greatly from a boosted decision tree,
which combines 20 kinematic variables to effectively suppress combinatorial
backgrounds. Charmless backgrounds are suppressed by exploiting the large
forward boost of the $D$ meson through a cut on its flight distance. The
signal yields are estimated by a simultaneous fit to 16 independent subsamples,
defined by the charges ($\times 2$), the $D$ final states ($\times 4$), and
the $K$ or $\pi$ nature of the bachelor hadron ($\times 2$).
Figures~\ref{fig:3-3} and~\ref{fig:3-4} show the projections of the $\pipi$ and
suppressed $\pipm\Kmp$ subsamples, respectively.
It is crucial to control the cross feed of the abundant $\Bm\to D\pim$ decays
into the signal decays. For this we rely on the two RICH detectors, which allow
to place particle identification cuts on the bachelor hadron. These cuts are $87.6\%$
efficient for kaons at a rate of $3.8\%$ misidentified pions.
Many systematic uncertainties cancel in the ratios Eqns.~\ref{eq:rkpi}-\ref{eq:rh}.
The remaining systematic uncertainties are dominated by an intrinsic charge
asymmetry of the detector, and by the uncertainty on the particle identification.
%
%
From the measured 13 observables the following established quantities can be deduced
(the full set is contained in our paper\cite{Aaij:2012kz}):
\begin{align*}
R_{\CP+} &= \phantom{-} 1.007 \pm 0.038 \pm 0.012~, \\
A_{\CP+} &= \phantom{-}  0.145 \pm 0.032 \pm 0.010~, \\
R_{K}^-  &= \phantom{-} 0.0073 \pm 0.0023 \pm 0.0004~, \\
R_{K}^+  &= \phantom{-} 0.0232 \pm 0.0034 \pm 0.0007~,
\end{align*}
where the first error is statistical and the second systematic; 
$R_{\CP+}$ is computed from $R_{\CP+} \approx \langle R_{K/\pi}^{KK} , R_{K/\pi}^{\pi\pi} \rangle / R_{K/\pi}^{K\pi}$
with an additional $1\%$ systematic uncertainty assigned to account for the
approximation; $A_{\CP+}$ is computed as $A_{\CP+} = \langle A_{K}^{KK} , A_{K}^{\pi\pi} \rangle$.
From the $R_{K}^\pm$ we also compute
\begin{align*}
  R_{\ads(K)}   &=       \phantom{-}  0.0152 \pm 0.0020 \pm 0.0004~, \\
  A_{\ads(K)}   &=       -0.52 \pm 0.15 \pm 0.02~,
\end{align*}
as $R_{\ads(K)} = (R_{K}^- + R_{K}^+) / 2$ and
$A_{\ads(K)} = (R_{K}^- - R_{K}^+) / (R_{K}^- + R_{K}^+)$.

To summarise, the $\Bpm\to D\Kpm$ \ads mode is observed with \mbox{$\approx10\sigma$} 
statistical significance when comparing the maximum likelihood to that of the null hypothesis.
This mode displays evidence ($4.0\sigma$) of a large negative asymmetry, consistent with previous 
experiments~\cite{Belle:2011ac,delAmoSanchez:2010dz,Aaltonen:2011uu}.
The combined asymmetry $A_{\CP+}$ is smaller than (but compatible with) previous
measurements~\cite{delAmoSanchez:2010ji,Aaltonen:2009hz}. It is $4.5\sigma$ significant.
We compare the maximum likelihood with that under the null-hypothesis in all 
three $D$ final states where the bachelor is a kaon, diluted by the non-negligible 
correlated systematic uncertainties. From this we observe, with a total significance 
of $5.8\sigma$, direct \CP violation in $\Bpm\to D\Kpm$ decays.



\section*{References}
\bibliographystyle{LHCb}
\bibliography{moriond}

\ifx\mcitethebibliography\mciteundefinedmacro
\PackageError{LHCb.bst}{mciteplus.sty has not been loaded}
{This bibstyle requires the use of the mciteplus package.}\fi
\providecommand{\href}[2]{#2}
\begin{mcitethebibliography}{10}
\mciteSetBstSublistMode{n}
\mciteSetBstMaxWidthForm{subitem}{\alph{mcitesubitemcount})}
\mciteSetBstSublistLabelBeginEnd{\mcitemaxwidthsubitemform\space}
{\relax}{\relax}

\bibitem{Aaij:2012jw}
LHCb collaboration, R.~Aaij {\em et~al.},
  \ifthenelse{\boolean{articletitles}}{{\it {Measurements of the branching
  fractions and CP asymmetries of $B^+ \to J/\psi \pi^+$ and $B^+ \to \psi(2S)
  \pi^+$ decays}}, }{}\href{http://arxiv.org/abs/1203.3592}{{\tt
  arXiv:1203.3592}}\relax
\mciteBstWouldAddEndPuncttrue
\mciteSetBstMidEndSepPunct{\mcitedefaultmidpunct}
{\mcitedefaultendpunct}{\mcitedefaultseppunct}\relax
\EndOfBibitem
\bibitem{Bhardwaj:2008ee}
Belle Collaboration, V.~Bhardwaj {\em et~al.},
  \ifthenelse{\boolean{articletitles}}{{\it {Observation of $B^\pm \to \psi(2S)
  \pi^\pm$ and search for direct CP-violation}},
  }{}\href{http://dx.doi.org/10.1103/PhysRevD.78.051104}{Phys.\ Rev.\  {\bf
  D78} (2008) 051104}, \href{http://arxiv.org/abs/0807.2170}{{\tt
  arXiv:0807.2170}}\relax
\mciteBstWouldAddEndPuncttrue
\mciteSetBstMidEndSepPunct{\mcitedefaultmidpunct}
{\mcitedefaultendpunct}{\mcitedefaultseppunct}\relax
\EndOfBibitem
\bibitem{PDG2010}
Particle Data Group, K.~Nakamura {\em et~al.},
  \ifthenelse{\boolean{articletitles}}{{\it {Review of particle physics}},
  }{}\href{http://dx.doi.org/10.1088/0954-3899/37/7A/075021}{J.\ Phys.\ G {\bf
  G37} (2010) 075021}\relax
\mciteBstWouldAddEndPuncttrue
\mciteSetBstMidEndSepPunct{\mcitedefaultmidpunct}
{\mcitedefaultendpunct}{\mcitedefaultseppunct}\relax
\EndOfBibitem
\bibitem{Aaij:2011in}
LHCb Collaboration, R.~Aaij {\em et~al.},
  \ifthenelse{\boolean{articletitles}}{{\it {Evidence for CP violation in
  time-integrated $D^0 \to h^-h^+$ decay rates}}, }{}Phys.\ Rev.\ Lett.\  {\bf
  108} (2012) 111602, \href{http://arxiv.org/abs/1112.0938}{{\tt
  arXiv:1112.0938}}\relax
\mciteBstWouldAddEndPuncttrue
\mciteSetBstMidEndSepPunct{\mcitedefaultmidpunct}
{\mcitedefaultendpunct}{\mcitedefaultseppunct}\relax
\EndOfBibitem
\bibitem{Aaij:2012qe}
LHCb collaboration, R.~Aaij {\em et~al.},
  \ifthenelse{\boolean{articletitles}}{{\it {First evidence of direct CP
  violation in charmless two-body decays of $B_s$ mesons}},
  }{}\href{http://arxiv.org/abs/1202.6251}{{\tt arXiv:1202.6251}}\relax
\mciteBstWouldAddEndPuncttrue
\mciteSetBstMidEndSepPunct{\mcitedefaultmidpunct}
{\mcitedefaultendpunct}{\mcitedefaultseppunct}\relax
\EndOfBibitem
\bibitem{HFAG}
Heavy Flavor Averaging Group, D.~Asner {\em et~al.},
  \ifthenelse{\boolean{articletitles}}{{\it {Averages of b-hadron, c-hadron,
  and tau-lepton properties}}, }{}\href{http://arxiv.org/abs/1010.1589}{{\tt
  arXiv:1010.1589}}, Updates available online at
  \url{http://www.slac.stanford.edu/xorg/hfag}\relax
\mciteBstWouldAddEndPuncttrue
\mciteSetBstMidEndSepPunct{\mcitedefaultmidpunct}
{\mcitedefaultendpunct}{\mcitedefaultseppunct}\relax
\EndOfBibitem
\bibitem{Aaltonen:2011qt}
CDF collaboration, T.~Aaltonen {\em et~al.},
  \ifthenelse{\boolean{articletitles}}{{\it {Measurements of direct CP
  violating asymmetries in charmless decays of strange bottom mesons and bottom
  baryons}}, }{}\href{http://dx.doi.org/10.1103/PhysRevLett.106.181802}{Phys.\
  Rev.\ Lett.\  {\bf 106} (2011) 181802},
  \href{http://arxiv.org/abs/1103.5762}{{\tt arXiv:1103.5762}}\relax
\mciteBstWouldAddEndPuncttrue
\mciteSetBstMidEndSepPunct{\mcitedefaultmidpunct}
{\mcitedefaultendpunct}{\mcitedefaultseppunct}\relax
\EndOfBibitem
\bibitem{Aaij:2012kz}
LHCb collaboration, R.~Aaij {\em et~al.},
  \ifthenelse{\boolean{articletitles}}{{\it {Observation of CP violation in
  $B^+ \to DK^+$ decays}}, }{}\href{http://arxiv.org/abs/1203.3662}{{\tt
  arXiv:1203.3662}}\relax
\mciteBstWouldAddEndPuncttrue
\mciteSetBstMidEndSepPunct{\mcitedefaultmidpunct}
{\mcitedefaultendpunct}{\mcitedefaultseppunct}\relax
\EndOfBibitem
\bibitem{Gronau:1990ra}
M.~Gronau and D.~London, \ifthenelse{\boolean{articletitles}}{{\it {How to
  determine all the angles of the unitarity triangle from $B_{d}^{0} \to D \KS$
  and $B_{s}^{0} \to D\phi$}},
  }{}\href{http://dx.doi.org/10.1016/0370-2693(91)91756-L}{Phys.\ Lett.\  {\bf
  B253} (1991) 483}\relax
\mciteBstWouldAddEndPuncttrue
\mciteSetBstMidEndSepPunct{\mcitedefaultmidpunct}
{\mcitedefaultendpunct}{\mcitedefaultseppunct}\relax
\EndOfBibitem
\bibitem{Gronau:1991dp}
M.~Gronau and D.~Wyler, \ifthenelse{\boolean{articletitles}}{{\it {On
  determining a weak phase from \CP asymmetries in charged \B decays}},
  }{}\href{http://dx.doi.org/10.1016/0370-2693(91)90034-N}{Phys.\ Lett.\  {\bf
  B265} (1991) 172}\relax
\mciteBstWouldAddEndPuncttrue
\mciteSetBstMidEndSepPunct{\mcitedefaultmidpunct}
{\mcitedefaultendpunct}{\mcitedefaultseppunct}\relax
\EndOfBibitem
\bibitem{Atwood:1996ci}
D.~Atwood, I.~Dunietz, and A.~Soni, \ifthenelse{\boolean{articletitles}}{{\it
  {Enhanced CP violation with $B \to K \Dz (\Dzb)$ modes and extraction of the
  CKM angle \g}},
  }{}\href{http://dx.doi.org/10.1103/PhysRevLett.78.3257}{Phys.\ Rev.\ Lett.\
  {\bf 78} (1997) 3257}, \href{http://arxiv.org/abs/hep-ph/9612433}{{\tt
  arXiv:hep-ph/9612433}}\relax
\mciteBstWouldAddEndPuncttrue
\mciteSetBstMidEndSepPunct{\mcitedefaultmidpunct}
{\mcitedefaultendpunct}{\mcitedefaultseppunct}\relax
\EndOfBibitem
\bibitem{Atwood:2000ck}
D.~Atwood, I.~Dunietz, and A.~Soni, \ifthenelse{\boolean{articletitles}}{{\it
  {Improved methods for observing CP violation in $\Bpm \to K D$ and measuring
  the CKM phase \g}},
  }{}\href{http://dx.doi.org/10.1103/PhysRevD.63.036005}{Phys.\ Rev.\  {\bf
  D63} (2001) 036005}, \href{http://arxiv.org/abs/hep-ph/0008090}{{\tt
  arXiv:hep-ph/0008090}}\relax
\mciteBstWouldAddEndPuncttrue
\mciteSetBstMidEndSepPunct{\mcitedefaultmidpunct}
{\mcitedefaultendpunct}{\mcitedefaultseppunct}\relax
\EndOfBibitem
\bibitem{Belle:2011ac}
Belle collaboration, Y.~Horii {\em et~al.},
  \ifthenelse{\boolean{articletitles}}{{\it {Evidence for the suppressed decay
  $\Bm \to D\Km, D \to \Kp\pim $}},
  }{}\href{http://dx.doi.org/10.1103/PhysRevLett.106.231803}{Phys.\ Rev.\
  Lett.\  {\bf 106} (2011) 231803}, \href{http://arxiv.org/abs/1103.5951}{{\tt
  arXiv:1103.5951}}\relax
\mciteBstWouldAddEndPuncttrue
\mciteSetBstMidEndSepPunct{\mcitedefaultmidpunct}
{\mcitedefaultendpunct}{\mcitedefaultseppunct}\relax
\EndOfBibitem
\bibitem{delAmoSanchez:2010dz}
Babar collaboration, P.~del Amo~Sanchez {\em et~al.},
  \ifthenelse{\boolean{articletitles}}{{\it {Search for $b \to u$ transitions
  in $\Bm \to D\Km$ and $\Bm \to D^*\Km$ decays}},
  }{}\href{http://dx.doi.org/10.1103/PhysRevD.82.072006}{Phys.\ Rev.\  {\bf
  D82} (2010) 072006}, \href{http://arxiv.org/abs/1006.4241}{{\tt
  arXiv:1006.4241}}\relax
\mciteBstWouldAddEndPuncttrue
\mciteSetBstMidEndSepPunct{\mcitedefaultmidpunct}
{\mcitedefaultendpunct}{\mcitedefaultseppunct}\relax
\EndOfBibitem
\bibitem{Aaltonen:2011uu}
CDF collaboration, T.~Aaltonen {\em et~al.},
  \ifthenelse{\boolean{articletitles}}{{\it {Measurements of branching fraction
  ratios and CP-asymmetries in suppressed $B^- \to D(\to K^+ \pi^-)K^-$ and
  $B^- \to D(\to K^+ \pi^-)\pi^-$ decays}},
  }{}\href{http://dx.doi.org/10.1103/PhysRevD.84.091504}{Phys.\ Rev.\  {\bf
  D84} (2011) 091504}, \href{http://arxiv.org/abs/1108.5765}{{\tt
  arXiv:1108.5765}}\relax
\mciteBstWouldAddEndPuncttrue
\mciteSetBstMidEndSepPunct{\mcitedefaultmidpunct}
{\mcitedefaultendpunct}{\mcitedefaultseppunct}\relax
\EndOfBibitem
\bibitem{delAmoSanchez:2010ji}
Babar collaboration, P.~del Amo~Sanchez {\em et~al.},
  \ifthenelse{\boolean{articletitles}}{{\it {Measurement of CP observables in
  $\Bpm \to D_{CP } \Kpm$ decays and constraints on the CKM angle \g}},
  }{}\href{http://dx.doi.org/10.1103/PhysRevD.82.072004}{Phys.\ Rev.\  {\bf
  D82} (2010) 072004}, \href{http://arxiv.org/abs/1007.0504}{{\tt
  arXiv:1007.0504}}\relax
\mciteBstWouldAddEndPuncttrue
\mciteSetBstMidEndSepPunct{\mcitedefaultmidpunct}
{\mcitedefaultendpunct}{\mcitedefaultseppunct}\relax
\EndOfBibitem
\bibitem{Aaltonen:2009hz}
CDF collaboration, T.~Aaltonen {\em et~al.},
  \ifthenelse{\boolean{articletitles}}{{\it {Measurements of branching fraction
  ratios and CP asymmetries in $\Bpm \to D_{CP} \Kpm$ decays in hadron
  collisions}}, }{}\href{http://dx.doi.org/10.1103/PhysRevD.81.031105}{Phys.\
  Rev.\  {\bf D81} (2010) 031105}, \href{http://arxiv.org/abs/0911.0425}{{\tt
  arXiv:0911.0425}}\relax
\mciteBstWouldAddEndPuncttrue
\mciteSetBstMidEndSepPunct{\mcitedefaultmidpunct}
{\mcitedefaultendpunct}{\mcitedefaultseppunct}\relax
\EndOfBibitem
\end{mcitethebibliography}

\end{document}